\documentclass[runningheads]{cl2emult}

\usepackage{makeidx}  
\usepackage{graphicx} 
\usepackage{subeqnar} 
\usepackage{multicol} 
\usepackage{cropmark} 
\usepackage{eso}      
\makeindex            

\newcommand{\lya}{Lyman-$\alpha$\ }
\newcommand{\magcir}{\ \raise -2.truept\hbox{\rlap{\hbox{$\sim$}}\raise5.truept
        \hbox{$>$}\ }}
\newcommand{\mincir}{\raise
-2.truept\hbox{\rlap{\hbox{$\sim$}}\raise5.truept
\hbox{$<$}\ }}

\begin{document}
\title*{CLUSTERING AT HIGH REDSHIFT}
\toctitle{ CLUSTERING AT HIGH REDSHIFT}
%
%
\titlerunning{Clustering at high-z}
%
\author{Stefano Cristiani\inst{1,2}
}
\authorrunning{Stefano Cristiani}
%
%
\institute{Space Telescope-European Coordinating Facility, European Southern Observatory
\and Dipartimento di Astronomia dell'Universit\`a di Padova}

\maketitle              

\begin{abstract}
Together with the CMB, the three sources of information that
astronomers have at high redshift as probes of the formation and
evolution of the LSS are QSOs, galaxies and absorbers observed in
the spectrum of distant background objects.
In this contribution I try to give 
a hint of historical perspective, following how
the technological advances have driven the emphasis from one class 
to another, in order to show what are the likely forthcoming
milestones.
\end{abstract}

\section{QSOs}
 
QSOs have been the first class of sources used to obtain direct
information about clustering at high redshift.
In the 80's they were the only available high-z objects bright 
enough to be discovered on photographic plates and observed in
relatively large quantities with
the existing spectrographic facilities.

Systematic searches began with the pioneering work of Osmer (1981) and
the detection of a clustering signal on scales $\sim 10$ Mpc
was achieved first using inhomogeneous QSO catalogs \cite{shaver84}
and then the statistically-well-defined samples used to study the QSO
luminosity function (LF) \cite{shanks87}.

QSOs display a number of appealing properties when compared to
galaxies as cosmological probes of the intermediate-linear regime of
clustering: they have a rather flat redshift distribution, their
point-like images are less prone to the surface-brightness biases
typical of galaxies and they sparse-sample the environment.
In recent times complete samples totaling about 2000 QSOs have been
assembled, providing a $\simeq 5 \sigma$ detection of the clustering
with an amplitude of $6 h^{-1}$ comoving Mpc 
\cite{andreani92,mo93,croom96},
consistent with or slightly larger than what is
observed for present-day galaxies and definitely lower than the
clustering of clusters.

The standard objection with regard to the information content in surveys of
QSOs and radio galaxies is that, since they are exotic objects, 
one risks learning 
possibly more about the formation and evolution of super-massive black holes 
in galactic nuclei, which is something interesting {\it per se}, than about
cosmology. 
In fact when the statistics on the clustering became sufficient to
address its evolution, and people started plotting amplitudes of the
correlation function (CF) as a function of redshift, surprisingly
the trends did not follow any of the canonical patterns of
constant, stable or collapsing clustering, in terms of the parameterization
$\xi(r,z) = \xi(r,z=0) (1+z)^{-(3+\epsilon-\gamma)}$,
where $\epsilon$ models the gravitational evolution of the structures.
\begin{figure}[t]
\centering
\includegraphics[width=.75\textwidth]{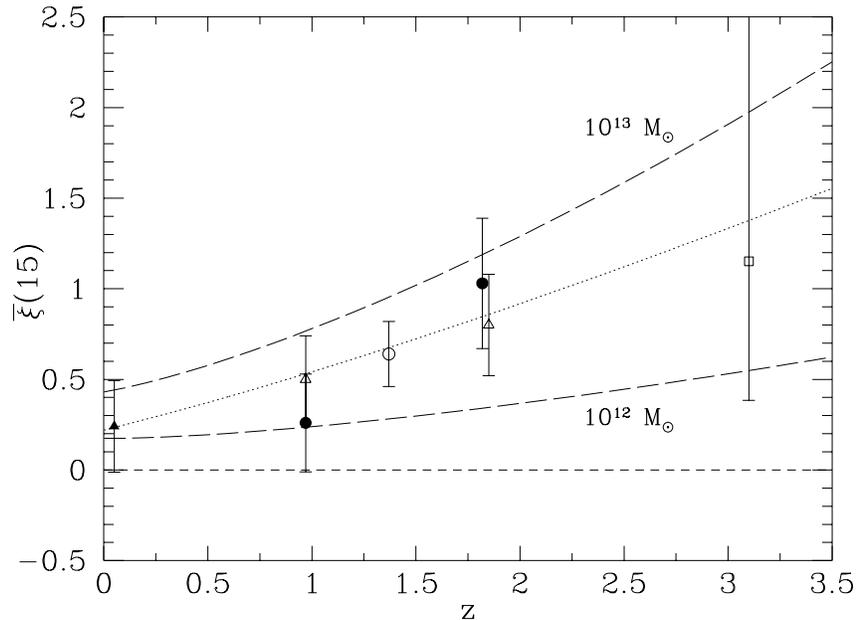}
\caption[]{
The amplitude of the average two-point correlation function (TPCF)
$\bar\xi(15~h^{-1}$ Mpc) as a
function of redshift (from La Franca, Andreani and Cristiani 1998).  
The dotted line is the $\epsilon=-2.5$ clustering evolution, 
the dashed lines show the evolution of the clustering for a minimum
halo mass of $10^{12}$ and $10^{13}$ $M_{\odot}$ $h^{-1}$
according to the transient model of Matarrese
et al. (1997).}
\label{QSOclustfig}
\end{figure}

Evidence is found for an {\it increase} of the clustering with
increasing redshift \cite{flf98}.  Not unreasonably, after all, since
the observed clustering is the result of an interplay between the
clustering of mass, decreasing with increasing redshift, and the bias,
increasing with redshift, convolved with the redshift distribution of
the objects \cite{mosca98,bagla98} and is
critically linked to the physics of the QSO formation and evolution.
Let us model, following  \cite{caval97},
the rise and fall of the QSO LF as the effect of
two components: the newly formed BH, which are dominant
at $z>3$ and the reactivated BH, which dominate at $z<3$.  The
reactivation is triggered by interactions taking place preferentially
in groups of galaxies.
In this way the clustering properties of QSOs are related to those of
transient, short-lived, objects \cite{matar97}. At high redshifts QSOs
correspond to larger (rarer) mass over-densities collapsing early and
cluster very strongly.  Then the clustering amplitude decreases until
the mass scale typical of a QSO reaches the value of the average
collapsing peaks, after which clustering may grow again.

If we think of QSOs as objects sparsely sampling halos with $M >
M_{\rm min}$, we can see from Fig.~\ref{QSOclustfig} that an $M_{\rm
min}= 10^{12} - 10^{13}~ h^{-1} M_{\odot}$ would provide at the same
time both the desired amount {\it and} evolution of clustering.
$5 \cdot 10^{12}$ M$_\odot$ is also the typical halo mass of the
groups of galaxies, in which interactions are most effective,
that fits correctly the evolution of the luminosity function.
 
The idea that QSOs reside in small to moderate groups of galaxies is
corroborated by the study of the environments of (radio-quiet) QSOs,
measuring QSO-galaxy clustering \cite{fisher96} or associated absorption
in close line-of-sight QSO pairs.
A recent result based on the WENSS and Green Bank radio surveys
\cite{rengelink99} also confirms the general tendency for radio sources
(QSOs+radio galaxies) to become increasingly biased tracers of the
matter distribution with increasing redshift.

The cosmogonic occurrence of the QSO phenomenon is beginning
to be understood. It is still arduous to turn to cosmology,
but it is also easy to predict that the SLOAN and 2DF QSO surveys
will bring dramatic progress in this direction.

\section{Clustering of Galaxies}

The type of rapid progress that in the meantime has taken place for
galaxies: CCDs have become larger and better, the HDF and other deep
fields have cleaned away some misconceptions about confusion
limits. Simple color criteria show remarkable efficiency, making it
possible to select high-z galaxies at an industrial rate.  It is
sufficient to obtain good multicolor imaging of deep fields down to $I
\simeq 25$ to get about one high-z galaxy per sq.arcmin.  The success
rate is so good that one can measure clustering at high-z almost
without spectroscopic redshifts.
The basic idea is that the redshifts estimated on the basis of the
photometry are sufficiently good to
subdivide the sample in redshift bins and compute the angular
correlation function (ACF) within the individual bins.
This is not the poor man's approach to the high-z universe,
but a technique which makes it possible to
derive redshifts (and clustering) for galaxies
which are about two magnitudes fainter than the deepest limits for
spectroscopic surveys (even with 10m class telescopes).
\begin{figure}
\centering
\includegraphics[width=.9\textwidth]{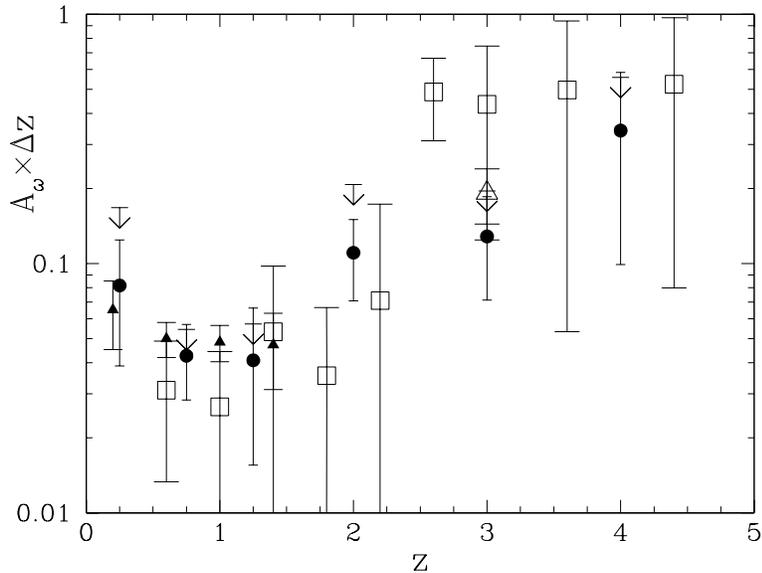}
\caption[]{
The amplitude of the angular correlation at 10 arcsec
 ($A_{\omega}$) as a function of redshift. 
Filled circles show the values obtained by Arnouts et al. 1999 (the
arrows are upper limits estimated on the basis of a contamination correction),
filled triangles the estimates obtained by
Connolly et al. (1998). The open triangle is the value for a
LBG sample of Giavalisco et al. (1998) and open squares refer to the
values obtained by Magliocchetti \& Maddox (1998). }
\label{fawz}
\end{figure}

We \cite{arnouts99} have built a photometric redshift code based on the
comparison of the observed colours of galaxies with those
expected from template SEDs derived from the GISSEL library
of models.
Spectroscopic redshifts are still important, to check that the code 
works well and that the estimated uncertainties 
correspond to the observed ones, with $\sigma_z$ increasing
from 0.1 to 0.2 with increasing redshift.
As a matter of fact it turns out that spectroscopic
redshifts are not always more reliable than the photometric ones:
in a few cases discrepancies have been found to be due to wrong 
spectroscopic estimates published in the literature.

Montecarlo simulations have been carried out
to infer the uncertainties also in the domain
inaccessible to spectroscopy ($I_{AB}\ge 26$)
and a comparison with other photometric methods has been performed to test the
stability as a function of the adopted templates.
This makes it possible to derive a map of the contamination effects
due to the uncertainties on the photometric redshifts 
and guided our choice of optimal redshift bins to minimize the 
effects of the errors, shot-noise and small field of view
(as a note in passing, also spectroscopic redshifts often end their lives 
in bins either to compute a LF or a global star formation rate or a CF).

The evolution of the ACF amplitude shows again
what is familiar from the case of QSOs: an initial decrease
followed by an increase.
As a consequence of the bias dependence on the redshift and on the
selection criteria of the samples, the behaviour of the galaxy
clustering cannot provide a straightforward indication of the
evolution of the underlying matter clustering.  For this reason, the
parametric form mentioned in Sect.1
cannot correctly describe the observations for any value of
$\epsilon$.
\begin{figure}
\centering
\includegraphics[width=1\textwidth]{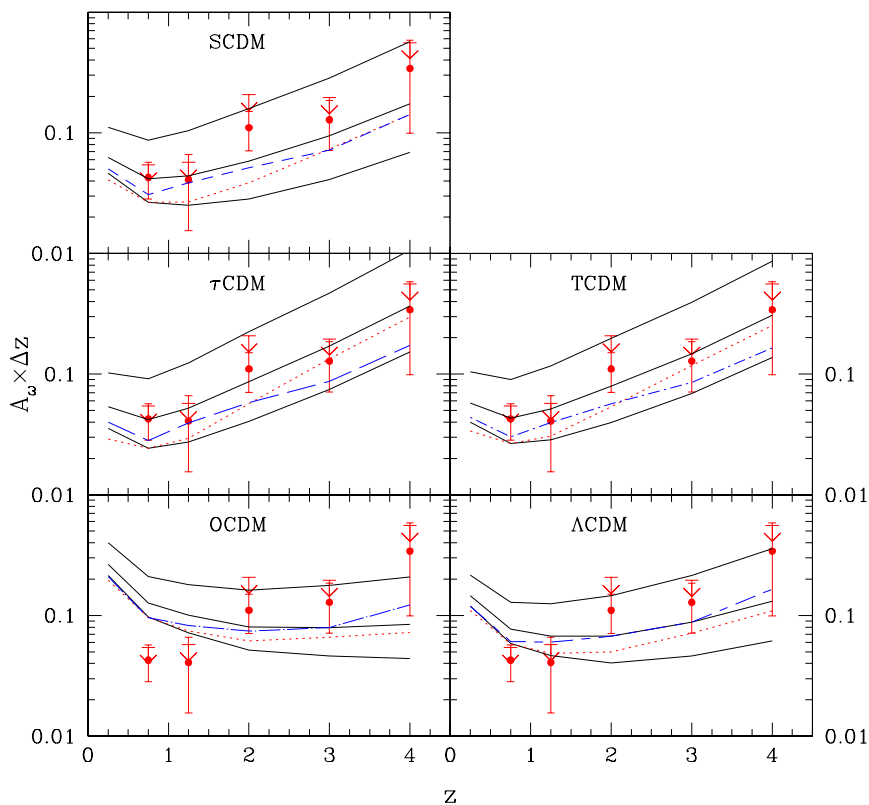}
\caption[]{
Comparison of the observed $A_{\omega}$ (filled
        circles for $I_{AB}\le 28.5$, arrows for the
        upper limits estimates) with the prediction of the
         various theoretical models. The solid lines show the measurements
        expected when a minimum mass $M_{\rm min} = 10^{10}$,
        $10^{11}$ and $10^{12} h^{-1} M_{\odot}$ is assumed; the lower
        curves refer to smaller masses. The dotted lines show the
        prediction obtained by using the median masses at any redshift
        estimated with GISSEL models. The
        dashed curves correspond to models where the masses necessary to
        reproduce the observed density of objects in each redshift bin
        are used. 
}
\label{fawmod} \end{figure}

We have compared our results with the theoretical predictions of a set
of different cosmological models belonging to the class of CDM
scenario.
The halo masses required to match the observations depend on the
adopted background cosmology. For Einstein de Sitter (EdS) universes,
the SCDM model reproduces the observed measurements if a typical
minimum mass of $ 10^{11} h^{-1}\ M_{\odot}$ is used, while the
$\tau$CDM and TCDM models require a lower typical mass of
$10^{10}-10^{10.5} h^{-1}\ M_{\odot}$.
For OCDM and $\Lambda$CDM
models, the mass is a function of redshift, with $M_{\rm min} \le
10^{10} h^{-1}\ M_{\odot}$ at $z\le 1$, $10^{11} h^{-1}\ M_{\odot}$
between $1\le z\le 3$ and $10^{12} h^{-1}\ M_{\odot}$ at $z\simeq 4$.
The higher masses required at high $z$ to reproduce the clustering
strength for these models are a consequence of the smaller bias they
predict at high redshift compared to the EdS models.
At $z \simeq 3$, the clustering strength and the
observed density of galaxies are in good agreement with the
theoretical predictions for any fashionable cosmological model.  At
$z\simeq 4$, the present analysis seems to be more discriminant.
Due to the remarkably high correlation strength,
for some models the observed density of galaxies starts to be
inconsistent with the required theoretical halo density.
The difference in the predicted masses
(a factor $\simeq$ 15 to 30 at $z\simeq$ 3 and 4) between EdS and non-EdS
universe models is also highly discriminant and
in principle testable in terms of measured
velocity dispersions.  
 
A prediction of the hierarchical models is the dependence of the
clustering strength on the limiting magnitude of the samples.  At
$z\simeq 3$ the density of galaxies in the HDF is approximately 65
times higher than the spectroscopic samples of LBGs
\cite{adelberger98,giavalisco98}.  Comparing our measurements with the
latter a clear trend of decreasing clustering strength with increasing
density is detected.  This result is in excellent agreement (both
qualitatively and quantitatively) with the predictions of hierarchical
models \cite{mo99} and, as noted by Adelberger et al. (1998),
suggests the existence of a strong relation between the halo mass and
the absolute UV luminosity: more massive haloes host the brighter
galaxies.
 
Finally, we have compared our $\sigma^{\rm gal}_8$ to that of the mass
predicted for three cosmologies to estimate the bias.  For all cases,
we found that the bias is an increasing function of redshift with
$b(z\simeq 0)\simeq 1 $ and $b(z\simeq 4) \simeq 4.5$ (for EdS
universes), and $b(z\simeq 0) \simeq 0.5 $ and $b(z\simeq 4)\simeq 3 $
(for open and $\Lambda$ universes).  This result confirms and extends
in redshift the results obtained for LBGs at $z\simeq 3$
\cite{adelberger98,giavalisco98}, suggesting that these high-redshift
galaxies are located preferentially in the rarer and denser peaks of
the underlying matter density field.

What are the descendants of the galaxies observed at $z\simeq
3-4$ in the HDF?
In a simple scenario assuming  that only one galaxy is hosted by the
descendants the resulting local bias implies that the descendants
in the case of OCDM and $\Lambda$CDM models can be
found among the brightest and more massive galaxies inside clusters,
while for EdS universes they are field or normal bright galaxies.

The present results have been obtained in a
relatively small field for which the effects of cosmic variance can be
important\cite{steidel98}. Nevertheless they show
a possibility of challenging cosmological parameters which becomes
particularly exciting in view of the rapidly growing wealth of
multi-wavelength photometric databases in various deep fields and
availability of 10m-class telescopes for spectroscopic follow-up in
the optical and near infrared.

\section{QSOs Strike Back: Clustering of Absorbers } 
After all galaxies are exotic objects too, in the sense that they
are also biased tracers of mass. There is, however, a way to follow
more ``normal'' matter at high redshifts: the absorption spectra of
QSOs, which sensitively probe the gaseous baryonic component of the
universe.  Sensitivity is the key word: with the present
instrumentation it is possible to detect neutral HI column densities
down to $10^{12}$ cm$^{-2}$, while for example 21 cm radio
observations which roughly correspond to the visible extent of a
galaxy are limited in the best cases to column densities that are 7
orders of magnitude larger.  In this way observations of the
Lyman-$\alpha$ forest reveal with an enormous dynamic range very
different structures, ranging from fluctuations of the diffuse
intergalactic medium to the interstellar medium in protogalactic
disks.

In recent times a large database of high-resolution ($\sim 10$ km/s),
high-S/N spectra of the Lyman forest has become available, allowing a
detailed investigation of the clustering of \lya lines, especially at high
redshift, where the density of lines provides particular sensitivity.
Cristiani et al (1997) have used more than 1600 \lya's
to detect a weak but significant signal, with $\xi\simeq 0.2$ on
scales of 100 km s$^{-1}$ at a $4.6 \sigma$ level.
Exploring the variations of the clustering as a function of the column
density a trend of increasing amplitude of the TPCF with increasing
column density is apparent, showing again that bias is at work. 

Metal systems show stronger clustering: the CIV CF at $z\simeq 2-2.5$
is consistent with a canonical power-law form with an $r_o = 3.5$ Mpc
and $\gamma \simeq 1.8$ with evidence for an increase
of the clustering strength with increasing column density
\cite{vale98a,quashnock98}.  The relatively large correlation scale
indicates that the CIV absorbers are biased tracers of relatively high
density regions.

To get a rough idea of the (over- and under-)densities corresponding
to \lya absorbers it is useful to estimate their size, on the basis of
the statistics of coincidences and anti-coincidences of absorptions in
close lines-of-sight to QSO pairs and groups \cite{vale98b}, 
from which one can infer
the ionization and the relationship between the observed optical
depths and the total density.  It turns out that these structures are
big (a few hundred kpc), highly ionized and at the lower column
densities ($\log N_{HI} \mincir 14$) probe weakly biased or even
anti-biased regions of the Universe.  It makes sense then to compute
the power spectrum (PS) of mass density fluctuations and assume that it
corresponds to the linear regime.  The recovery of the 3-D PS is a
complex procedure and various recipes have been proposed by a number
of authors \cite{croft98,hui99,nusser99}.
The present results indicate that the PS amplitude is
consistent with some scale-invariant, COBE-normalized CDM models (an
open model with $\Omega_0=0.4$ OCDM, variants of the SCDM) and
inconsistent with others (the CCDM model, with $\Omega_0=1$, $h=0.5$,
$\sigma_8=1.2$)\cite{croft98}.  
Even with limited dynamic range and substantial
statistical uncertainty, a measurement of the PS that has no unknown
``bias factors'' offers many opportunities for testing theories of
structure formation and constraining cosmological parameters.

In a similar fashion, it has been proposed to apply the
Alcock-Paczinsky test to QSO groups to investigate the cosmological
geometry \cite{hui99}. This test, based on the comparison between the
clustering properties of the (Lyman-$\alpha$) absorbers, namely the
TPCF, observed along the line of sight and the corresponding estimate
in the transverse direction, is especially sensitive to
$\Omega_{\Lambda}$ and should be able to discriminate between an
($\Omega_m = 0.3,~ \Lambda=0$) and an ($\Omega_m = 0.7,~ \Lambda=0.3$)
universe at a $4 \sigma$ level by observing $\sim 25$ QSO pairs.
 
Future prospects look really exciting.  Starting at the end of
1999, the echelle spectrograph UVES will be available on UT2 of the
VLT, extending, with respect to HIRES at Keck, the possibility of
high-resolution ($ \Re \sim 50000$) 
observations down to at least $V\sim 20$ and to a
shorter wavelength range (i.e. to more numerous, lower-redshift
QSOs).  Then (by 2001) the FLAMES facility (see {\tt
http://http.hq.eso.org/instruments/flames/}) will allow observation
with UVES of up to 8 QSOs at the same time in a field of 25 arcmin
diameter and/or 135 targets in the same area with an intermediate/high
resolution spectrograph, GIRAFFE. 
The Alcock-Paczinsky test and other cosmologically discriminant
measures could be carried out in a few nights.
\section{Acknowledgments}
I wish to thank P.Andreani, S.Arnouts, V.D'Odorico, S.D'Odorico, A.Fontana,
E.Giallongo, F. La Franca, F.Lucchin, S.Matarrese, L.Moscardini with
whom I have been collaborating and P.Bristow for
carefully reading the manuscript.

\clearpage
\addcontentsline{toc}{section}{Index}
\flushbottom
\printindex


\begin{thebibliography}{99}
%
\addcontentsline{toc}{section}{References}

\bibitem{adelberger98} Adelberger K.L., Steidel C.C., Giavalisco M.,
Dickinson M.E., Pettini M., Kellogg M., 1998, ApJ {\bf 505}, 18
\bibitem{andreani92} Andreani P., Cristiani S. 1992, ApJ {\bf 398}, L13
\bibitem{arnouts99} Arnouts S., Cristiani S., Moscardini L.,
Matarrese L., Lucchin F., Fontana A., Giallongo E. astro-ph/9902290 
\bibitem{bagla98} Bagla J.S., 1998, MNRAS {\bf 297}, 251
\bibitem{caval97} Cavaliere A., Perri M., Vittorini V., 1997,
Mem.S.A.It. {\bf 68}, 27
\bibitem{connolly98} Connolly A.J., Szalay A.S., Brunner R.J., 1998,
ApJ, 499, L125
\bibitem{croft98} Croft R.A.C., et al., 1998, astro-ph/9809401
\bibitem{croom96} Croom S.M., Shanks T., 1996, MNRAS {\bf 281}, 89
\bibitem{vale98a} D'Odorico V., Cristiani S., D'Odorico S., Fontana A.,
Giallongo E., 1998a, A\&AS {\bf 127} 217
\bibitem{vale98b} D'Odorico V., Cristiani S., D'Odorico S., Fontana A.,
Giallongo E., Shaver P., 1998b, A\&A {\bf 339}, 678
\bibitem{fisher96} Fisher K.B., Bahcall J.N., Kirhakos S., Schneider
D. P., 1996, ApJ {\bf 468}, 469
\bibitem{giavalisco98} Giavalisco M., Steidel C.C., Adelberger K.L.,
Dickinson M.E., Pettini M., Kellogg M., 1998, ApJ {\bf 503}, 543
\bibitem{hui99}
Hui L., Stebbins A., Burles S., 1999, ApJ {\bf 511}, L5
\bibitem{flf98} La Franca F., Andreani P., Cristiani S. 1998, ApJ
{\bf 497}, 529
\bibitem{maglio99} Magliocchetti M., Maddox S.J., 1998, astro-ph/9811320
\bibitem{matar97} Matarrese S., Coles P., Lucchin F., Moscardini, L.,
1997, MNRAS {\bf 286}, 115 
\bibitem{mo93} Mo H.J., Fang L.-Z., 1993, ApJ {\bf 410}, 493
\bibitem{mo99} Mo H.J., Mao S., White S.D.M., 1999, MNRAS {\bf 304}, 175
\bibitem{mosca98} Moscardini L., Coles P., Lucchin F., 
Matarrese S., 1998, MNRAS {\bf 299}, 95
\bibitem{nusser99} Nusser A., Haehnelt M., 1999, MNRAS {\bf 303}, 179
\bibitem{quashnock98}
Quashnock J.M., Vanden Berk D.E., 1998, ApJ {\bf 500}, 28
\bibitem{rengelink99} Rengelink R., 1999, Ph.D. Thesis
\bibitem{shanks87} Shanks T., Fong R., Boyle B., Peterson B.A., 
1987, MNRAS {\bf 227}, 739
\bibitem{shaver84} Shaver P.A., 1984, A\&A {\bf 136}, L9
\bibitem{steidel98} Steidel C., 1998, astro-ph/9811400
\end{thebibliography}
\end{document}